\documentclass[aps,twocolumn,showpacs,epsf]{revtex4}
\usepackage{graphicx}

\def\nbZ{{\mathchoice {\hbox{$\sf\textstyle Z\kern-0.4em Z$}}
{\hbox{$\sf\textstyle Z\kern-0.4em Z$}} {\hbox{$\sf\scriptstyle
Z\kern-0.3em Z$}}  {\hbox{$\sf\scriptscriptstyle Z\kern-0.2em Z$}}}}

\begin{document}

\title{Quasiperiodic tilings under magnetic field}
\author{Julien Vidal}
\email{vidal@gps.jussieu.fr}
\author{R\'emy Mosseri}
\email{mosseri@gps.jussieu.fr}

\affiliation{
Groupe de Physique des Solides, CNRS UMR 7588, Universit\'{e}s
Paris 6 et  Paris 7, 2, place Jussieu, 75251 Paris Cedex 05 France}

\begin{abstract}

We study the electronic properties of a two-dimensional quasiperiodic tiling, the isometric generalized
Rauzy tiling, embedded in a magnetic field. Its energy spectrum is computed in a tight-binding approach
by means of the recursion method. Then, we study the quantum dynamics of wave packets and discuss the
influence of the magnetic field on the diffusion and spectral exponents.  Finally, we consider a
quasiperiodic superconducting wire network with the same geometry and we determine the critical
temperature as a function of the magnetic field.

\end{abstract}


\maketitle

%
%
%%%%%%%%%%%%%%%%%%%%%%%%%%%%
\section{Introduction}
%%%%%%%%%%%%%%%%%%%%%%%%%%%%
%
%
For periodic systems such as the square, the triangular or the hexagonal lattice embedded in
a magnetic field, it is well-known that commensurability effects between the characteristic
magnetic length and the lattice period may lead to very interesting features. For example, if we
consider a standard tight-binding model, the energy spectrum displays a complex multifractal
structure when varying the magnetic flux per unit cell as illustrated by the famous Hofstadter
butterfly \cite{Hofstadter,Claro_Wannier,Rammal_hexa}. However, this spectrum is strongly
geometry-dependent since even for periodic tilings, some anomalies can appear leading to interesting 
quantum interference phenomena \cite{Vidal_Cages,Vidal_Cages_big}.
This dependence has led several groups to analyze the interplay between the quasiperiodicity and the
magnetic frustration \cite{Arai,Hatakeyama1,Hatakeyama2,Schwabe_Mag_field}. 
Nevertheless, in these studies, it is difficult to know the real contribution of the quasiperiodic
order since the incommensurability between the tiles area is another source of aperiodicity in the
spectrum \cite{Hatakeyama1}. We must indeed distinguish several classes of tilings depending on their
structural order (periodic, quasiperiodic or disordered), their topology (trivial or nontrivial) and
the ratio of the tiles area (rational or irrational).  If we focus on quasiperiodically ordered
structures, there is actually one type of tiling (those of codimension one) that has never been studied
so far which consists in a nontrivial topology (several types of coordination numbers) with
commensurate tile areas.

The aim of this work is to study such tilings by considering a modified version of the
two-dimensional generalized Rauzy tilings (GRT) presented in Ref. \cite{Vidal_Rauzy} that are built
with tiles having the same areas. Note that a great advantage of such systems is that all the
physical quantities are periodic functions of the magnetic field so that we can restrict our analysis to
a finite range of flux values. \\

This paper is organized as follows:
in the next section, we briefly present the two-dimensional isometric generalized Rauzy tilings
(iGRT) whose properties will always be compared to the square lattice ones. In Sect. \ref{Butterflies},
we introduce the tight-binding Hamiltonian and we compute the energy spectrum of both tilings.   We
show that despite the quasiperiodic order, the  spectrum of the iGRT has a fine nontrivial
structure. To get more precise informations on the spectrum and the eigenstates, we compute in
Sect. \ref{Dynamics} the time evolution of wave packets for different fluxes per plaquette $\phi$
and we focus on the average autocorrelation function and the mean square spreading. For the
iGRT a singular continuous spectrum and a sub-ballistic propagation is found for rational and 
irrational $f=\phi/\phi_0$ where $\phi_0$ is the flux quantum. By contrast, for the square lattice,
the propagation is  ballistic and the spectrum absolutely continuous for rational $f$, whereas the
spectrum is singular continuous and the spreading sub-ballistic for irrational $f$.
Sect. \ref{Superconducting} is devoted to the superconducting wire network with the iGRT
geometry.  After mapping the linearized Ginzburg-Landau equations onto a tight-binding model with
nonconstant hopping terms, we compute the transition temperature as a function of the magnetic field
and compare it with  those of the square lattice. A cusp-like structure is clearly observed and shows
the importance of the quasiperiodic order in this context.

%
%
%%%%%%%%%%%%%%%%%%%%%%%%%%%%%%%%%%%%%%%%%%
\section{The isometric generalized Rauzy tilings}
\label{IGRT}
%%%%%%%%%%%%%%%%%%%%%%%%%%%%%%%%%%%%%%%%%%
%
%
The GRTs are codimension one quasiperiodic structures that can easily be
built in any dimension $D$ using the standard cut and project method \cite{Vidal_Rauzy}. These
tilings have a complex topological structure with sites of coordination number ranging from $D+1$
to $2D+1$. In this study, we focus on the two-dimensional GRT whose contruction is based on the 
irrational solution $\theta\simeq 1.839$ of the equation:
%
%
%%%%%%%%%%%%%%%%%
\begin{equation}
x^3=x^2+x+1
\mbox{.}
\end{equation}
%%%%%%%%%%%%%%%%%
%
%
Contrary to previous studies \cite{Vidal_ICQ7,Triozon_Rauzy,Jagannathan_Rauzy}
concerning these tilings, we consider here an isometric version of the $2D$ GRT where all the
edge lengths are equal. This can be very easily done by projecting the selected sites of
the $3D$ cubic lattice perpendicularly to the direction $(1,1,1)$ instead of the Rauzy
direction $(1,\theta,\theta^2)$. Such a choice is motivated by the fact that since these
tilings are codimension one tilings, equal edge lengths imply equal tile areas. This property
which is only true for codimension one tilings 
\footnote{For example, the Penrose (codimension 3) and Octagonal (codimension 2) tilings
have incommensurate tile areas but equal edge lengths.} allows us to concentrate on the role
played by the topological quasiperiodicity related to the connectivity matrix.
%
%
%%%%%%%%%%%%%%%%%%
\begin{figure}[ht]
\includegraphics[width=60mm]{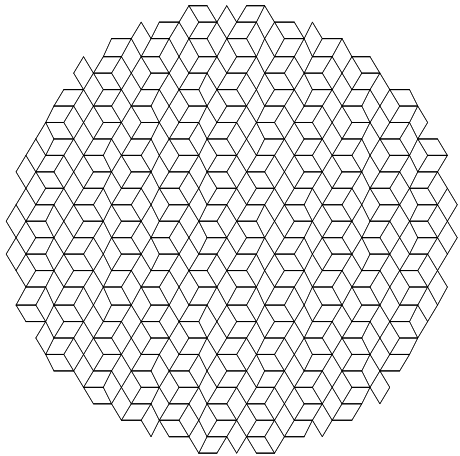}
\caption{ 
A piece of the $2D$ isometric generalized Rauzy tiling. All the tiles have the same area.
}
\label{tilings}
\end{figure}
%%%%%%%%%%%%%%%%%%
%
%

%
%
%%%%%%%%%%%%%%%%%%%%%%%%%%%%%%%%%%%%%%%%%%
\section{Hamiltonian and butterflies}
\label{Butterflies}
%%%%%%%%%%%%%%%%%%%%%%%%%%%%%%%%%%%%%%%%%%
%
%
In this study, we consider a standard tight-binding Hamiltonian given by:
%
%
%%%%%%%%%%%%%%%%%
\begin{equation}
H = -\sum_{\langle  i,j\rangle} t_{ij}  \, |i \rangle \langle  j|
\label{Hamiltonian}
\mbox{,}
\end{equation}
%%%%%%%%%%%%%%%%%
%
%
where $|i\rangle$ is a localized orbital on site $i$.
The hopping term $t_{ij}=1$ if $i$ and $j$ are nearest neighbors and
$0$ otherwise. In the presence of a magnetic field \cite{Peierls}, $t_{ij}$ is multiplied
by a phase factor $e^{i \gamma _{ij}}$ involving the vector potential ${\bf {A}}$:
%
%
%%%%%%%%%%%%%%%%%%%%%%%
\begin{equation}
\gamma _{ij}={\frac{2\pi }{\phi _{0}}}\int_{i}^{j}{\bf A}.d {\bf l}
\label{defgamma}
\mbox{,}
\end{equation}
%%%%%%%%%%%%%%%%%%%%%%%
%
%
where $\phi _{0}=hc/e$ is the flux quantum. In the following, we only consider the case
of a uniform magnetic field ${\bf B}=B{\bf z}$ which can be obtained, for example, with the
Landau gauge ${\bf A}=B(0,x,0)$, and we denote by \mbox{$\phi = B a^{2}\sqrt{3}/2$} the magnetic 
flux through an elementary tile. It is also useful to introduce the dimensionless parameter
$f=\phi/ \phi_0$ known as the reduced flux. 

The study of the zero-field spectrum of $H$ has been presented in 
Ref.~\cite{Vidal_ICQ7,Triozon_Rauzy,Jagannathan_Rauzy} for the nonisometric GRTs, but for $B=0$
the tile areas is an irrelevant parameter since $H$ is, in this case,  simply proportionnal to the
connectivity matrix. The spectral measure is absolutely continuous, the spectrum is gapless, and the
weakly critical eigenstates, are responsible for an anomalous diffusion.
To obtain the spectrum of $H$ for $B\neq 0$, we have used the recursion method \cite{Haydock_rec3}
also known as Lanczos algorithm \cite{Paige}. This powerful tool allows one to compute the local
density of states (LDOS) of a given initial state. The main advantage of this method is that we can
easily compute this LDOS  for very large system size (about $ 10^6$ sites), by diagonalizing only a
small half-chain of a few hundred sites. However, it does just give a local information but, if one
considers an initial random phase state, its LDOS gives a fairly good approximation of the total
density of states. 
To compute these spectra we have computed 500 recursion coefficients in a circular cluster of
785757 sites for the square lattice, and 460 coefficients in a circular cluster of 767604 sites for
the iGRT. 
We must also mention an important point concerning the edge states. Indeed, since we compute the
recursion coefficients up to a given order, there are obviously some spurious states that
arise due to the fact that the recursion cluster is finite with open boundary conditions. To get rid
of them, we have selected eigenvectors for which the amplitude on the initial state of the recursion
was greater than a given threshold (here $10^{-9}$).

As it can be seen in Fig. \ref{Butterflies:fig}, the spectrum of the square lattice obtained here
with the recursion scheme is in very good agreement with those computed by Hofstadter in 1976 using
the mappping onto the Harper equation \cite{Hofstadter}. As it can be readily seen, both spectra
display the symmetry $E
\leftrightarrow -E$ related to the fact that both tilings are bipartite. In addition, since there is
only one characteristic area, the Hamiltonian is a periodic function of $f$ with period 1. Finally, the
symmetry with respect to $f=1/2$ just reflects the symmetry $t_{ij}\leftrightarrow - t_{ij}$. 
At small fields, the emergence of Landau levels, expected in the
continuum limit, is clearly observed. In the square lattice, it is well known that rational values of
$f=p/q$ lead to an absolutely continuous spectrum made up of $q$ subbands that can touch each
other as for $f=1/2$. Of course, nothing similar is expected for quasiperiodic tilings since
no periodicity exists.Nevertheless, it is interesting to note that for the iGRT,
there is still a complex gap structure with notably a large gap that can be followed
continuously from $f=0$ (where it vanishes)  till $f=1/2$.
We can also see that the iGRT spectrum display a fine subband structure that would
require a detailed but difficult analyzis since even in the zero field case, it is nontrivial.

%
%
%%%%%%%%%%%%%%%%%%
\begin{figure}[ht] 
\includegraphics[width=3.7in]{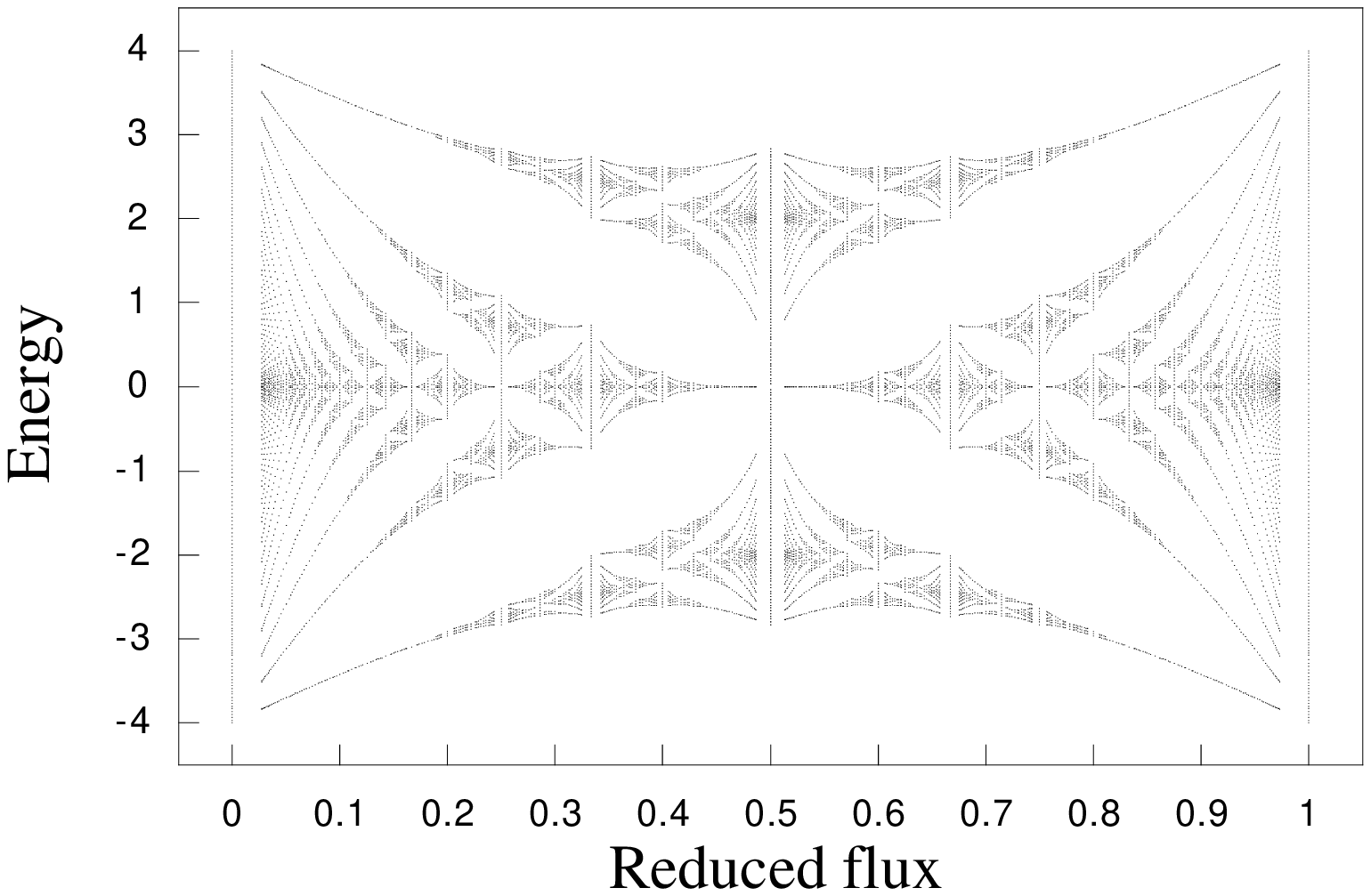}
\end{figure}
\begin{figure}[ht] 
\vspace{-12mm}
\includegraphics[width=3.7in]{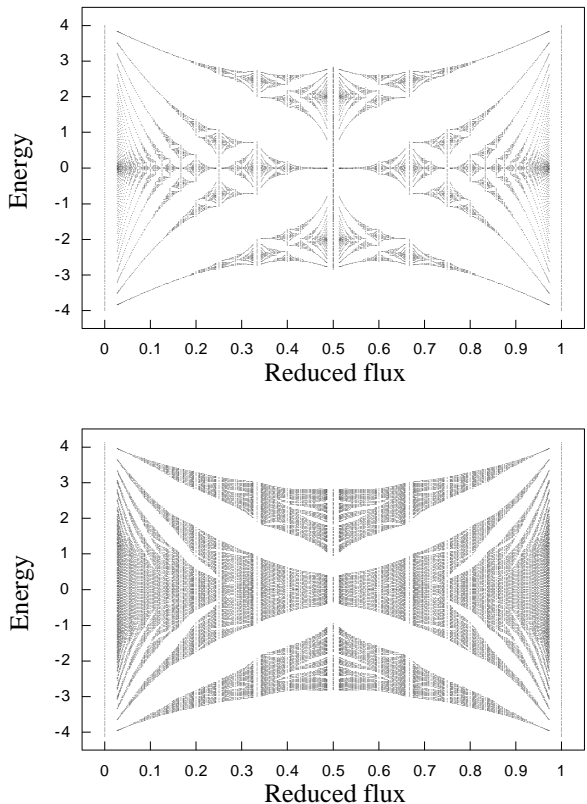}
\vspace{-15mm}
\caption{
Energy spectrum as function of the reduced flux $f=\phi/ \phi_0$ for the square lattice (top) and for
the iGRT (bottom). For convenience, we have only considered rational values of  $f=p/q$
for $q \leq 37$.}
\label{Butterflies:fig}
\end{figure}
%%%%%%%%%%%%%%%%%%
%
%

To go beyond this basic description, we have analyzed the nature of the spectrum
for particular values of the reduced flux by studying the quantum dynamics of wave packets
in both structures for remarkable values of the reduced flux.

%
%
%%%%%%%%%%%%%%%%%%%%%%%%%%%%%%%%%%%%%%%%%%
\section{Wave packets dynamics}
\label{Dynamics}
%%%%%%%%%%%%%%%%%%%%%%%%%%%%%%%%%%%%%%%%%%
%
%
The quantum dynamics in two-dimensional quasiperiodic tilings has been widely investigated within
the last ten years either by exact diagonalizations \cite{Passaro_Octo,Zhong,Yuan} or by
approximating the evolution operator $e^{-iHt}$ \cite{Triozon_Rauzy,Vidal_ICQ7}.
Here we have used the Second Order Differencing Scheme \cite{Numerique}, that consists, for a given
initial state $|\psi(0)\rangle$, to make the following approximation:
%
%
%%%%%%%%%%%%%%%%%%%%%%%
\begin{equation}
|\psi(t+\Delta t)\rangle =|\psi(t-\Delta t)\rangle -2i \Delta t H |\psi(t)\rangle
\mbox{.}
\end{equation}
%%%%%%%%%%%%%%%%%%%%%%%
%
%
Contrary to exact diagonalizations, this method allows one to investigate the dynamics on very
large systems (about one million sites). Practically, choosing a time step $\Delta t=0.05$ is
sufficient to get a good accuracy and all our results have been obtained with this value.  
We have considered two quantities of interest: the averaged autocorrelation function
%
%
%%%%%%%%%%%%%%%%%%%%%%%
\begin{equation}
C(t)={1\over t} \int_0^{t} dt' P(t') \sim t^{-\alpha}
\mbox{,}
\end{equation}
%%%%%%%%%%%%%%%%%%%%%%%
%
%
where $P(t)=|\langle \psi(0)|\psi(t)\rangle|^2$ and the mean square spreading 
%
%
%%%%%%%%%%%%%%%%%%%%%%%
\begin{equation}
\Delta R^2(t)=
\langle \psi(t)|\hat R^2|\psi(t)\rangle - (\langle \psi(t)|\hat R|\psi(t)\rangle)^2
\sim t^{2\beta}
\mbox{,}
\end{equation}
%%%%%%%%%%%%%%%%%%%%%%%
%
%
where $\hat R$ is the position operator, the origin being  taken at the center of the cluster.  
The spectral and diffusion exponents $\alpha$ and $\beta$ are defined from the long time behavior of
$C$ and $\Delta R^2$ respectively. If $\alpha=1$, the spectrum is absolutely continuous, if
$\alpha=0$, the spectrum is pure point, and if $0<\alpha<1$ the spectrum is singular
continuous. Concerning the diffusion exponent, $\beta=1$ corresponds to a ballistic
propagation whereas $0<\beta<1$ defines a sub-ballistic regime ($\beta=1/2$ is the diffusive case). 

For the square lattice, since all sites are equivalent, we have considered an initial state
localized on a single (central) site. 
For the iGRT, the situation is more complicated since, as for the Octagonal
tiling \cite{Passaro_Octo}, the exponents depends, a priori on
the initial conditions as shown in Ref.~\cite{Triozon_Rauzy,Vidal_ICQ7}. 
Of course, this is only true at short times, since in the asymptotic regime,  we expect
$\alpha=\inf_E\{\alpha(E)\}$, $\beta=\sup_E\{\beta(E)\}$ for $E$ in the spectrum.  
Here, we have built an initial random phase state localized on a circular cluster ($\sim 10^2$ sites)
whose radius is about 1/100 of the cluster  considered for the computation, and we have checked that
our results were weakly sensitive to the precise random phase configuration chosen.
However, we can never ensure that we have reached the asymptotic regime so that the exponents given
here just give a qualitative idea	of the wave packet spreadings. 

We have considered several special values of the
reduced flux: (i) $f=0$ for which the band width are maximum for both lattices~; (ii) $f=1/2$  for
which the band width of the iGRT is minimum~; (iii) $f=\sqrt{2}$ for which the band width of the
square lattice is minimum~; (iv) $f=\theta$ which is a natural irrational number in
this context. 
Our results are plotted in Fig.\ref{Exposants:fig1} and  Fig.\ref{Exposants:fig2}.

For the square lattice, everything is qualitatively known for these values. Indeed, when $f$
is a rational number, the eigenstates are Bloch waves, the spectrum is absolutely continuous
($\alpha=1$) and the spreading is ballistic ($\beta=1$). For $f=0$, when starting from a single
site, one can even easily show that: $P(t)=J^4_0(2 t)$ which behaves as $t^{-2}$ at large time  
\footnote{$J_0$ denotes the Bessel function of order $0$.} so that $C(t) \sim t^{-1}$ and $\Delta
R^2(t)=4 t^2$.  For irrational $f$, the spectrum is known to be singular continuous
\cite{Hofstadter,KPG,Zhong}, and the propagation sub-ballistic
\cite{Piechon,Ketzmerick}. Nevertherless, we emphasize that existing results about the dynamics have
been obtained by analyzing the Harper chain and do not reflect the dynamics on the square
lattice.  Indeed, the mapping of the square lattice onto the Harper chain results from  a
decoupling of the eigenstates given by
$\Psi(x,y)=e^{i k_y y} \psi(x)$ which is due the special Landau gauge choice. Thus, the dynamics
discussed in Refs. \cite{KPG,Zhong,Piechon,Ketzmerick} only concerns the $x$ direction (at fixed $k_y$)
and thus do not give the exponent of the two-dimensional square lattice. For example, we have computed
the dynamics for $f=(1+\sqrt{5})/2=\tau$ and we have found $\alpha=0.50(1)$ in the square lattice
instead of $\alpha=0.14(1)$ in the Harper chain \cite{KPG,Zhong}. 

%
%
%%%%%%%%%%%%%%%%%%
\begin{figure}[ht] 
\includegraphics[width=3.7in]{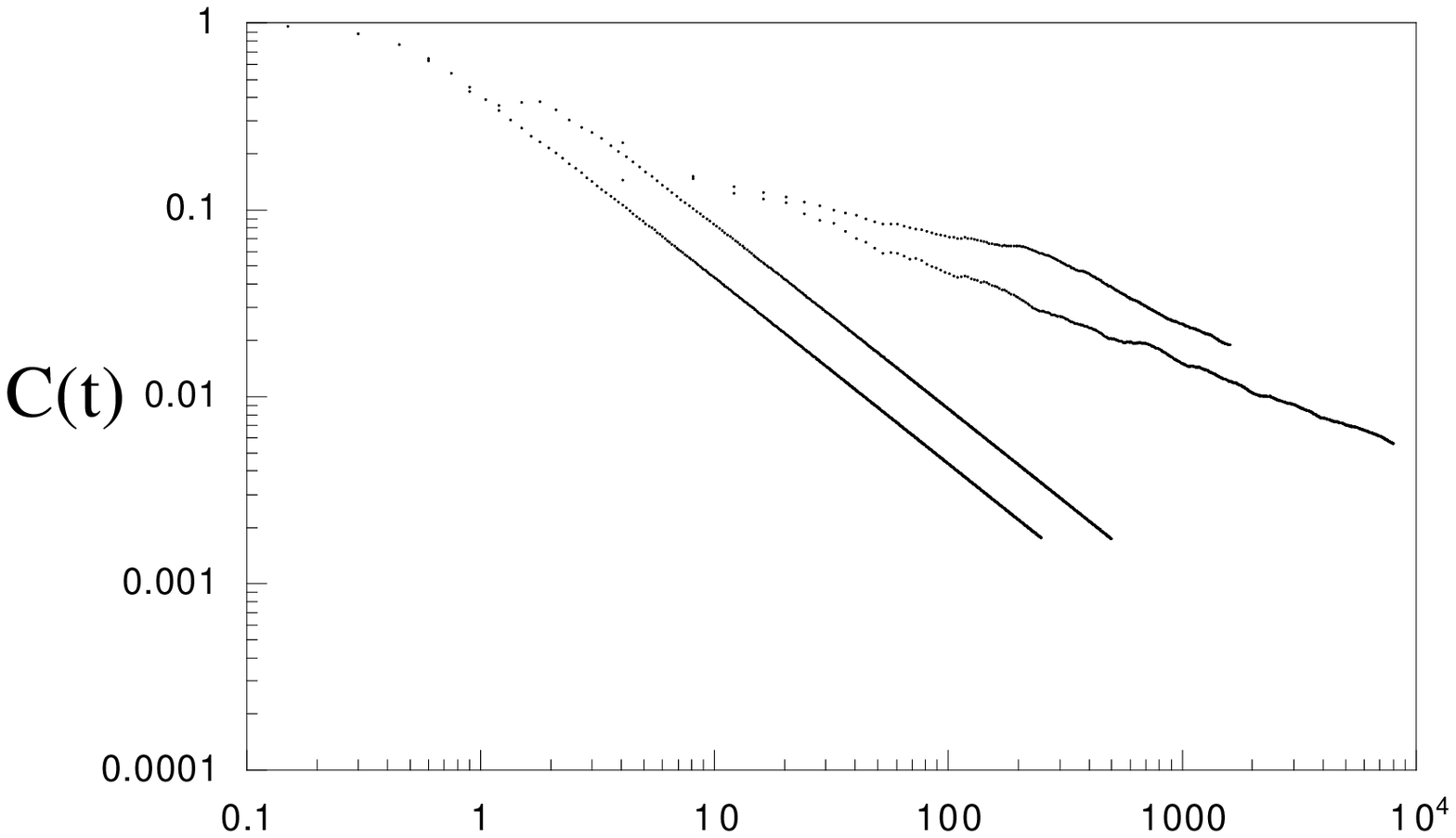}
\end{figure}
\begin{figure}[ht]
\vspace{-20mm}
\includegraphics[width=3.7in]{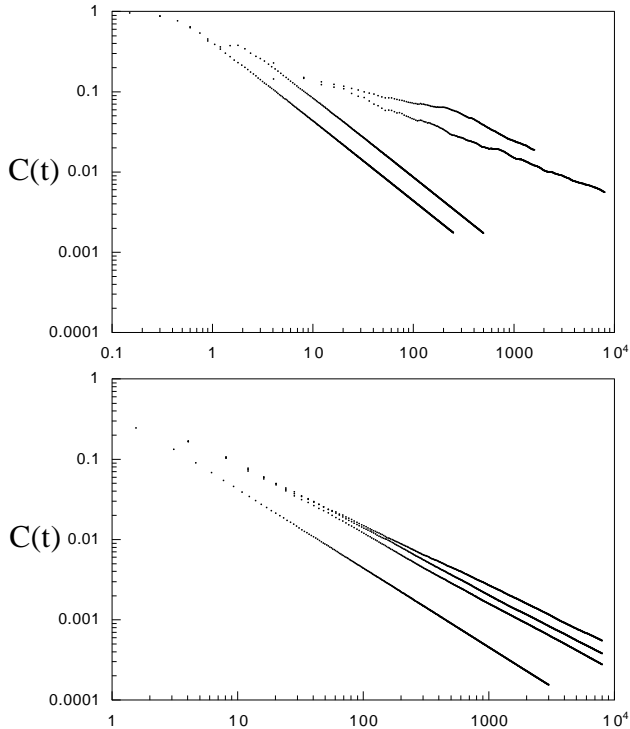}
\vspace{-17mm}
\caption{
Averaged autocorrelation function. Top: square lattice with $f=0, 1/2, \sqrt{2}, \theta$ (from
bottom to top). Bottom: iGRT for $f=0, 1/2, \theta, \sqrt{2}$ (from bottom to top).}
\label{Exposants:fig1}
\end{figure}
%%%%%%%%%%%%%%%%%%
%
%

%\vspace{-113mm}
%\hspace{-4mm}
%{\large $C(t)$}
%\vspace{113mm}

%\vspace{-58mm}
%\hspace{-4mm}
%{\large $C(t)$}
%\vspace{50mm}

Interestingly, for $f=\sqrt{2}$, we have found a diffusive propagation with $\beta=0.50(1)$ whereas for
$f=\theta$, the problem is more complicated since there are clearly two regimes. If we consider only $t
\gtrsim 250$ we obtain,
\mbox{$\beta=0.94(2)$} (weakly sub-ballistic motion) but we cannot decide whether it is
an intermediate regime or not.  The exponents obtained by fitting our data are given in \mbox{Table
I}.

For the iGRT, the situation is different since even for $f=0$, the spectrum has some critical states
responsible for a sub-ballistic spreading \cite{Vidal_Rauzy}. However, as shown
in Ref.~\cite{Triozon_Rauzy} the diffusion exponent is strongly energy-dependent and might even be
very close to 1 at the band edges. 
In the asymtotic regime, a quasi-ballistic propagation should therefore be recovered. Contrary to the
square lattice, a sub-ballistic spreading is found for $f=1/2$ with
$\beta=0.61(1)$ which is not really surprising since even though $1/2$ is a rational number, the
underlying quasiperiodicity of the iGRT does not allow for the contruction extended eigenstates. In a
sense, there are no commensurability effects for rational $f$. 
%
%
%%%%%%%%%%%%%%%%%%
\begin{figure}[ht] 
\includegraphics[width=3.7in]{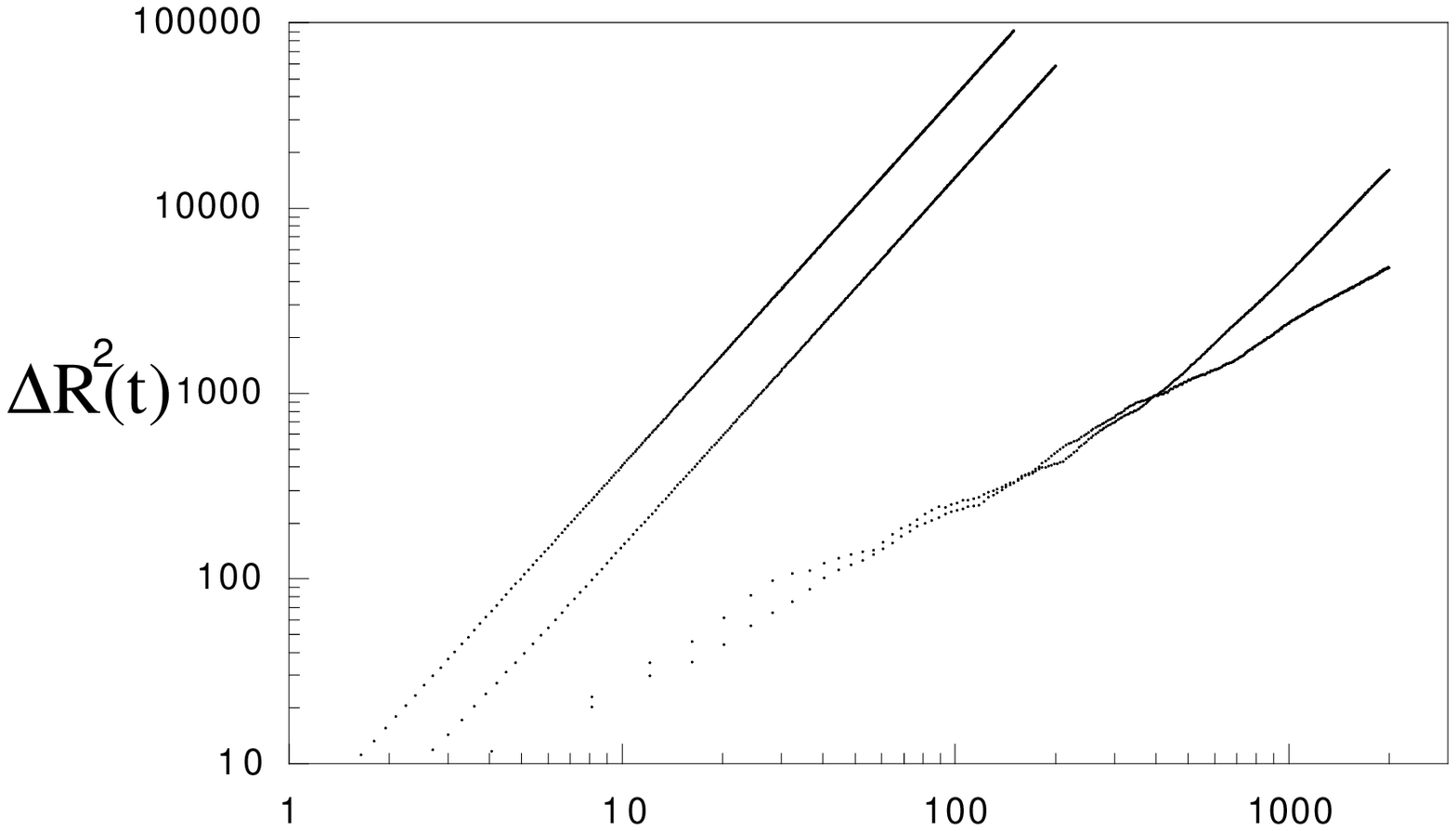}
\end{figure}
\begin{figure}[ht]
\vspace{-20mm}
\includegraphics[width=3.7in]{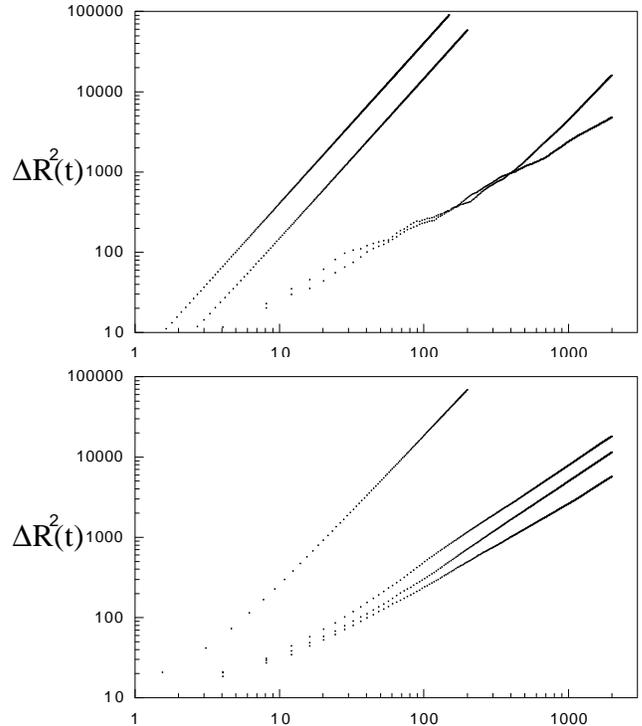}
\vspace{-17mm}
\caption{
Mean square spreading. Top: square lattice for \\$f=0, 1/2,
\theta, \sqrt{2},$ (from top to bottom).
Bottom: iGRT for $f=0, 1/2, \theta, \sqrt{2}$ (from
top to bottom).}
\label{Exposants:fig2}
\end{figure}
%%%%%%%%%%%%%%%%%%
%
%
%
%
%%%%%%%%%%%%%%%%%%%%%%%%%%%%%%%%%%%%%%%%%%%%%%%%
\begin{table}[ht]
\begin{center}
\begin{tabular}{|c|c|c|c|c|c|}
\hline 
               & $f=0$       &   $f=1/2$     & $f=\sqrt{2}$  & $f=\theta$  \\
\hline
square lattice &             &               &               &             \\
$\alpha$       &   0.99(1)   &   0.99(1)     &   0.47(3)     &  0.55(5)    \\
$\beta$        &   0.99(1)   &   0.99(1)     &   0.50(1)     &  0.94(2)    \\
\hline
iGRT           &             &               &               &             \\
$\alpha$       &   0.98(1)   &   0.84(1)     &   0.77(1)     &  0.80(1)    \\
$\beta$        &   0.94(2)   &   0.61(1)     &   0.57(2)     &  0.60(1)    \\
\hline
\end{tabular}
\end{center}
\caption{Spectral and diffusion exponents for the square lattice and for the iGRT.}
\end{table}
%%%%%%%%%%%%%%%%%%%%%%%%%%%%%%%%%%%%%%%%%%%%%%%%
%
%
For $f=\sqrt{2}$, it is
interesting to note that the propagation is faster in the iGRT ($\beta=0.57(2)$) than in the
square lattice $\beta=0.50(1)$. This means that the quasiperiodic order can break the
destructive quantum interference effects that slow the spreading down for incommensurate $f$ in
periodic structures. Nevertheless, this is not always the case, since 
for $f=\theta$, the propagation is faster in the square lattice than in the iGRT. 
Concerning the nature of the spectrum, it is always found to be clearly singular continuous but for
$f=0$, we cannot be sure that $\alpha\neq 1$ in the asymptotic regime.

%
%
%%%%%%%%%%%%%%%%%%%%%%%%%%%%%%%%%%%%%%%%%%
\section{Superconducting wire networks}
\label{Superconducting}
%%%%%%%%%%%%%%%%%%%%%%%%%%%%%%%%%%%%%%%%%%
%
%
In this section, we analyze the role of the quasiperiodic order on the superconducting-normal
transition temperature as a function of the magnetic field $T_c(f)$. As explained above, the main
advantage of the iGRT is that all its elementary tiles have the same area. Thus, we can
really analyze the role of the topological quasiperiodic order which has, in fact, never been
investigated. 

The study of quasiperiodic wire networks  has been initially motivated by several 
experiments \cite{Behrooz_PRL,Behrooz_PRB,Wang2,Springer} in which the superconducting
transition temperature as a function of the magnetic field was determined.
Using a mapping between the linearized Ginzburg-Landau equations and the tight-binding model,  the rich
structure of this phase boundaries has been then computed by Nori and coworkers \cite{Nori1,Nori3}. 
The main idea of this correspondence developped by Alexander \cite{Alexander_fils_supra} and
De Gennes \cite{DeGennes_fils_supra} is to integrate the LGL equations on each wire and to
obtain a set of equations that determines the order parameter
$\psi_i$ at each nodes $i$. This can be simply achieved using the continuity of the order parameter and
the current conservation at each node.  Denoting by $l_{ij}$ the distance between the (connected) nodes
$i$ and $j$ and by $\xi$ the coherence length, the current conservation at each node $i$ expresses as:
%
%
%%%%%%%%%%%%%%%%%%%%%%%
\begin{equation}
-\psi_i \sum_{\langle i,j\rangle} \cot(l_{ij}/\xi)+ \sum_{\langle i,j\rangle}
\psi_j e^{i \gamma_{ij}} /\sin(l_{ij}/\xi)=0
\label{LGL}
\mbox{,}
\end{equation}
%%%%%%%%%%%%%%%%%%%%%%%
%
%
where the elementary flux quantum involved on the definition (\ref{defgamma}) of $\gamma_{ij}$
is here $\phi'_0=hc/2e$. The coherence length depends on the temperature:
%
%
%%%%%%%%%%%%%%%%%%%%%%%
\begin{equation}
\xi(T)=\xi(0) \left({T_c(0) \over T_c(0)-T} \right)^{1/2}
\mbox{,}
\end{equation}
%%%%%%%%%%%%%%%%%%%%%%%
%
%
where $T_c(0)$ is the transition temperature at zero field. For a given value of the magnetic field,
the transition temperature is thus given by the highest temperature for which a nontrivial solution 
of (\ref{LGL}) exists.
 If we now consider a system where all the lengths are equal $l_{ij}=l$, and if we set
$\phi_i=\sqrt{z_i} \psi_i$ where $z_i$ is the coordination of the node $i$, Eq. (\ref{LGL}) can be
rewritten as \cite{Nori3}:
%
%
%%%%%%%%%%%%%%%%%%%%%%%
\begin{equation}
\cos(l/\xi) \phi_i = \sum_{\langle i,j\rangle} (z_i z_j)^{-1/2}  e^{i \gamma_{ij}} \phi_j
\label{LGL2}
\mbox{.}
\end{equation}
%%%%%%%%%%%%%%%%%%%%%%%
%
%
Setting with $t_{ij}=-(z_i z_j)^{-1/2}$ and $E=\cos(l/\xi)$, we obtain the secular system given
by the Hamiltonian (\ref{Hamiltonian}) and thus the field dependence of the transition
temperature:
%
%
%%%%%%%%%%%%%%%%%%%%%%%
\begin{equation}
1-\frac{T_{c}(f)}{T_{c}(0)}=\frac{\xi(0)^{2}}{l^{2}}
\arccos^2(E_0(f))
\label{Tc-vs-energy}
\end{equation}
%%%%%%%%%%%%%%%%%%%%%%%
%
%
where $E_0(f)$ is the highest eigenvalue of this modified Hamiltonian for a reduced flux 
$f=2 \pi \phi/\phi'_0$.

For the square lattice, the first experiments have been performed by Pannetier {\it et
al.} \cite{Pannetier2} and have confirmed the validity of these predictions. Indeed, a self-similar
cusp-like structure for rational $f$ has been clearly identified in agreement with the fractal pattern
of $E_0(f)$ already pointed by Hofstadter \cite{Hofstadter}.  For quasiperiodic systems, experiments by
Behrooz and coworkers \cite{Behrooz_PRL,Behrooz_PRB} have also revealed an cusp-like structure that
can be understood in terms of incommensurability between the tiles area \cite{Nori1}. 
%
%
%%%%%%%%%%%%%%%%%%
\begin{figure}[ht] 
\includegraphics[width=3.7in]{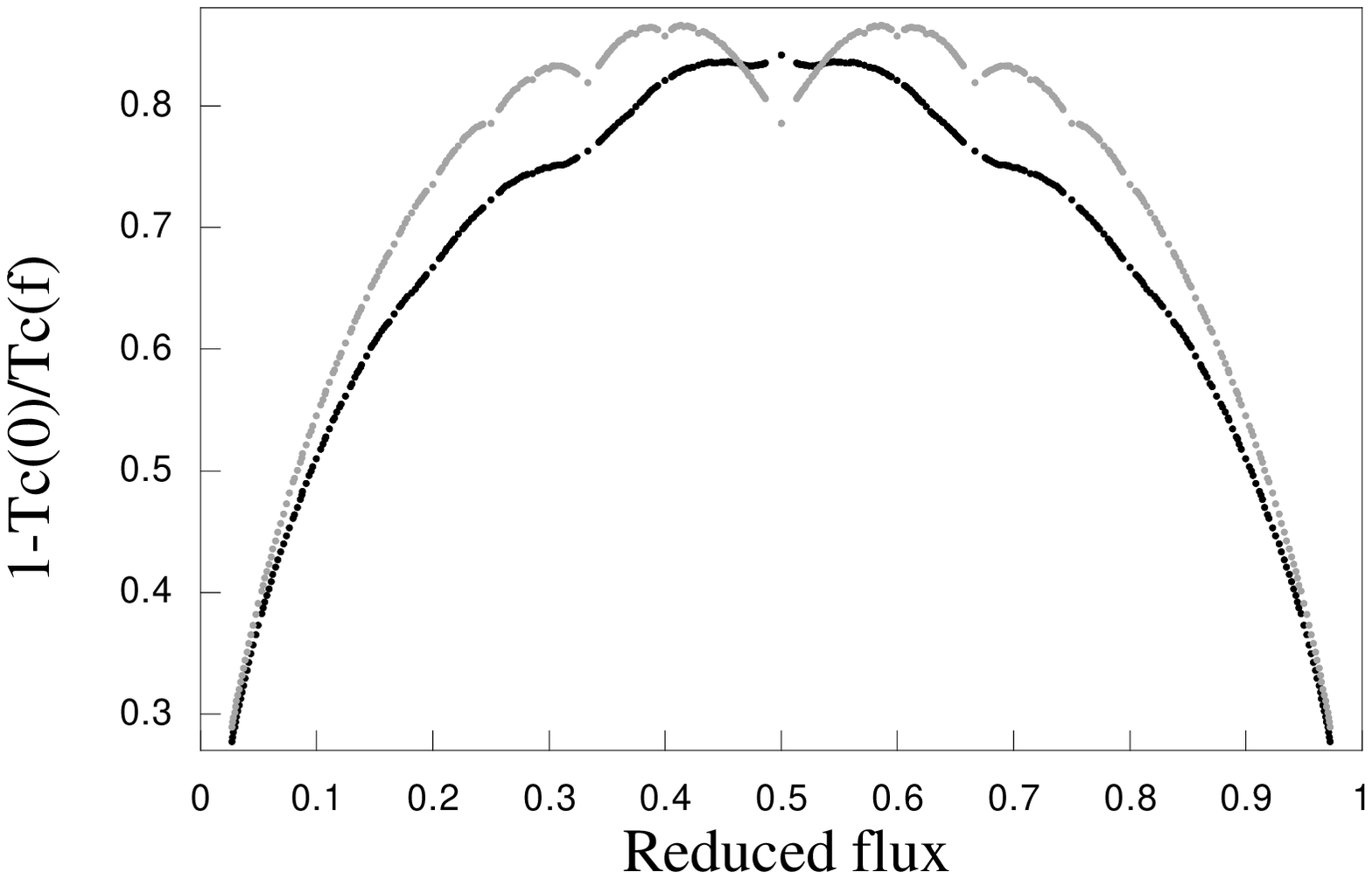}
\end{figure}
\begin{figure}[ht]
%\hspace{mm}
\vspace{-51.7mm}
\hspace{14.85mm}
\includegraphics[width=38mm]{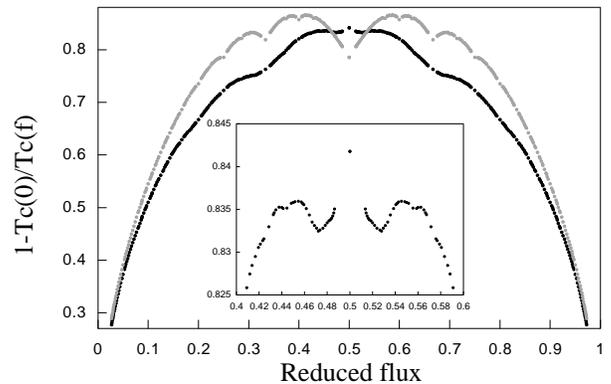}
\vspace{8mm}
\caption{ 
Reduced critical temperature as a function of the reduced flux $f$ for the square lattice (grey
curve) and for the iGRT (black curve). Inset: zoom around $f=1/2$ for the iGRT.
}
\label{Tc}
\end{figure}
%%%%%%%%%%%%%%%%%%
%
%

We have computed $T_c(f)$ for the square lattice and for the iGRT using, as previously, the
recursion method. As explained in Ref.~\cite{Nori3}, this is a particularly well-adapted algorithm to
obtain the ground state of the modified Hamiltonian which is in this case equal to $-E_0(f)$.  
The results displayed in Fig. \ref{Tc}, clearly show
the strong cusp-like structure for the square lattice but nothing similar seems to occur for
iGRT. However, when zooming, we can distinguish a fine structure in $T_c(f)$ but with oscillations much
smaller than those of the square lattice.  We also note that, for the iGRT, the minimum of $T_c(f)$ is
obtained for $f=1/2$ where an inverse-cusp is observed as for the Dice lattice \cite{Abilio_T3} or for
the Kagome lattice \cite{Xiao_Kagome,Lin_Nori}. This rich cusp structure is rather strange for this
tiling since there is no incommensurability effect due to either the ratio of the tile areas (which are
all equal for the iGRT) or to the lattice periodicity.

%
%
%%%%%%%%%%%%%%%%%%%%%%%%%%%%%%%%%%%%%%%%%%
\section{Conclusion}
We have studied the spectrum and the quantum dynamics of an electron in a quasiperiodic tiling embedded
in a magnetic field. As in periodic structures, its spectrum displays a complex energy pattern when the
magnetic flux is varied. We have shown that for several values of the reduced flux $f$, the wave
packet spreading is sub-ballistic with a field-dependent diffusion exponent.  For $f=\sqrt{2}$, we have
even found a propagation which is faster in the quasiperiodic tiling than in the square lattice for
which the motion is, in this case, diffusive.  It would be very interesting to
study the influence of the disorder (topological, geometrical, or Anderson-like) on these exponents
since, even in the zero-field case, nontrivial behaviours have already been observed
\cite{Roche_Review}.

We have also determined the superconducting-normal transition line $T_c(f)$ that displays a cusp-like
structure revealing the importance of the quasiperiodic order.  However, the analysis of the cusp
positions clearly requires further investigations. It would be also important to study the transition
line for codimension one random tilings that also have one type of tile but no quasiperiodic
order. In this case, the absence of cusp would confirm the particular role played by the 
quasiperiodicity. \\

We would like to thank Cl. Aslangul and B. Dou\c{c}ot for fruitful discussions.

\label{Conclusion}
%%%%%%%%%%%%%%%%%%%%%%%%%%%%%%%%%%%%%%%%%%
%
%

%\bibliography{bibliotheque}

\begin{thebibliography}{10}
\expandafter\ifx\csname bibnamefont\endcsname\relax
  \def\bibnamefont#1{#1}\fi
\expandafter\ifx\csname bibfnamefont\endcsname\relax
  \def\bibfnamefont#1{#1}\fi
\expandafter\ifx\csname url\endcsname\relax
  \def\url#1{\texttt{#1}}\fi
\expandafter\ifx\csname urlprefix\endcsname\relax\def\urlprefix{URL }\fi
\providecommand{\bibinfo}[2]{#2}
\providecommand{\eprint}[2][]{\url{#2}}

\bibitem{Hofstadter}
\bibinfo{author}{\bibfnamefont{D.}~\bibnamefont{Hofstadter}},
  \bibinfo{journal}{Phys. Rev. B} \textbf{\bibinfo{volume}{14}},
  \bibinfo{pages}{2239} (\bibinfo{year}{1976}).

\bibitem{Claro_Wannier}
\bibinfo{author}{\bibfnamefont{F.~H.} \bibnamefont{Claro}} \bibnamefont{and}
  \bibinfo{author}{\bibfnamefont{G.~H.} \bibnamefont{Wannier}},
  \bibinfo{journal}{Phys. Rev. B} \textbf{\bibinfo{volume}{19}},
  \bibinfo{pages}{6068} (\bibinfo{year}{1979}).

\bibitem{Rammal_hexa}
\bibinfo{author}{\bibfnamefont{R.}~\bibnamefont{Rammal}}, \bibinfo{journal}{J.
  Phys. (Paris)} \textbf{\bibinfo{volume}{46}}, \bibinfo{pages}{1345}
  (\bibinfo{year}{1985}).

\bibitem{Vidal_Cages}
\bibinfo{author}{\bibfnamefont{J.}~\bibnamefont{Vidal}},
  \bibinfo{author}{\bibfnamefont{R.}~\bibnamefont{Mosseri}}, \bibnamefont{and}
  \bibinfo{author}{\bibfnamefont{B.}~\bibnamefont{Dou\c{c}ot}},
  \bibinfo{journal}{Phys. Rev. Lett.} \textbf{\bibinfo{volume}{81}},
  \bibinfo{pages}{5888} (\bibinfo{year}{1998}).

\bibitem{Vidal_Cages_big}
\bibinfo{author}{\bibfnamefont{J.}~\bibnamefont{Vidal}},
  \bibinfo{author}{\bibfnamefont{P.}~\bibnamefont{Butaud}},
  \bibinfo{author}{\bibfnamefont{B.}~\bibnamefont{Dou\c{c}ot}},
  \bibnamefont{and} \bibinfo{author}{\bibfnamefont{R.}~\bibnamefont{Mosseri}},
  \bibinfo{journal}{Phys. Rev. B} \textbf{\bibinfo{volume}{64}},
  \bibinfo{pages}{155306} (\bibinfo{year}{2001}).

\bibitem{Arai}
\bibinfo{author}{\bibfnamefont{M.}~\bibnamefont{Arai}},
  \bibinfo{author}{\bibfnamefont{T.}~\bibnamefont{Tokihiro}}, \bibnamefont{and}
  \bibinfo{author}{\bibfnamefont{T.}~\bibnamefont{Fujiwara}},
  \bibinfo{journal}{J. Phys. Soc. Jpn.} \textbf{\bibinfo{volume}{56}},
  \bibinfo{pages}{1642} (\bibinfo{year}{1987}).

\bibitem{Schwabe_Mag_field}
\bibinfo{author}{\bibfnamefont{H.}~\bibnamefont{Schwabe}},
  \bibinfo{author}{\bibfnamefont{G.}~\bibnamefont{Kasner}}, \bibnamefont{and}
  \bibinfo{author}{\bibfnamefont{H.}~\bibnamefont{B\"{o}ttger}},
  \bibinfo{journal}{Phys. Rev. B} \textbf{\bibinfo{volume}{56}},
  \bibinfo{pages}{8026} (\bibinfo{year}{1997}).

\bibitem{Hatakeyama1}
\bibinfo{author}{\bibfnamefont{T.}~\bibnamefont{Hatakeyama}} \bibnamefont{and}
  \bibinfo{author}{\bibfnamefont{H.}~\bibnamefont{Kamimura}},
  \bibinfo{journal}{Solid State Commun.} \textbf{\bibinfo{volume}{62}},
  \bibinfo{pages}{79} (\bibinfo{year}{1987}).

\bibitem{Hatakeyama2}
\bibinfo{author}{\bibfnamefont{T.}~\bibnamefont{Hatakeyama}} \bibnamefont{and}
  \bibinfo{author}{\bibfnamefont{H.}~\bibnamefont{Kamimura}},
  \bibinfo{journal}{J. Phys. Soc. Jpn.} \textbf{\bibinfo{volume}{58}},
  \bibinfo{pages}{260} (\bibinfo{year}{1989}).

\bibitem{Vidal_Rauzy}
\bibinfo{author}{\bibfnamefont{J.}~\bibnamefont{Vidal}} \bibnamefont{and}
  \bibinfo{author}{\bibfnamefont{R.}~\bibnamefont{Mosseri}},
  \bibinfo{journal}{J. Phys. A} \textbf{\bibinfo{volume}{34}},
  \bibinfo{pages}{3927} (\bibinfo{year}{2001}).

\bibitem{Vidal_ICQ7}
\bibinfo{author}{\bibfnamefont{J.}~\bibnamefont{Vidal}} \bibnamefont{and}
  \bibinfo{author}{\bibfnamefont{R.}~\bibnamefont{Mosseri}}, in
  \emph{\bibinfo{booktitle}{Proceedings of the 7th International Conference on
  Quasicrystals}}, edited by
  \bibinfo{editor}{\bibfnamefont{F.}~\bibnamefont{G\"ahler}},
  \bibinfo{editor}{\bibfnamefont{P.}~\bibnamefont{Kramer}},
  \bibinfo{editor}{\bibfnamefont{H.~R.} \bibnamefont{Trebin}},
  \bibnamefont{and} \bibinfo{editor}{\bibfnamefont{K.}~\bibnamefont{Urban}}
  (\bibinfo{publisher}{Elsevier}, \bibinfo{address}{Switzerland},
  \bibinfo{year}{2000}), vol. \bibinfo{volume}{A294-A296}, p.
  \bibinfo{pages}{572}.

\bibitem{Triozon_Rauzy}
\bibinfo{author}{\bibfnamefont{F.}~\bibnamefont{Triozon}},
  \bibinfo{author}{\bibfnamefont{J.}~\bibnamefont{Vidal}},
  \bibinfo{author}{\bibfnamefont{R.}~\bibnamefont{Mosseri}}, \bibnamefont{and}
  \bibinfo{author}{\bibfnamefont{D.}~\bibnamefont{Mayou}},
  \bibinfo{journal}{Phys. Rev. B} \textbf{\bibinfo{volume}{65}},
  \bibinfo{pages}{220202} (\bibinfo{year}{2002}).

\bibitem{Jagannathan_Rauzy}
\bibinfo{author}{\bibfnamefont{A.}~\bibnamefont{Jagannathan}},
  \bibinfo{journal}{Phys. Rev. B} \textbf{\bibinfo{volume}{64}},
  \bibinfo{pages}{140201} (\bibinfo{year}{2001}).

\bibitem{Peierls}
\bibinfo{author}{\bibfnamefont{R.~E.} \bibnamefont{Peierls}},
  \bibinfo{journal}{Z. Phys.} \textbf{\bibinfo{volume}{80}},
  \bibinfo{pages}{763} (\bibinfo{year}{1933}).

\bibitem{Haydock_rec3}
\bibinfo{author}{\bibfnamefont{R.}~\bibnamefont{Haydock}},
  \emph{\bibinfo{title}{Solid State Physics}} (\bibinfo{publisher}{Academic
  Press}, \bibinfo{address}{New York}, \bibinfo{year}{1980}),
  chap.~\bibinfo{chapter}{35}, p. \bibinfo{pages}{216}.

\bibitem{Paige}
\bibinfo{author}{\bibfnamefont{C.}~\bibnamefont{Paige}}, \bibinfo{journal}{J.
  Inst. Math. Appl.} \textbf{\bibinfo{volume}{10}}, \bibinfo{pages}{373}
  (\bibinfo{year}{1978}).

\bibitem{Passaro_Octo}
\bibinfo{author}{\bibfnamefont{B.}~\bibnamefont{Passaro}},
  \bibinfo{author}{\bibfnamefont{C.}~\bibnamefont{Sire}}, \bibnamefont{and}
  \bibinfo{author}{\bibfnamefont{V.~G.} \bibnamefont{Benza}},
  \bibinfo{journal}{Phys. Rev. B} \textbf{\bibinfo{volume}{46}},
  \bibinfo{pages}{13751} (\bibinfo{year}{1992}).

\bibitem{Zhong}
\bibinfo{author}{\bibfnamefont{J.~X.} \bibnamefont{Zhong}} \bibnamefont{and}
  \bibinfo{author}{\bibfnamefont{R.}~\bibnamefont{Mosseri}},
  \bibinfo{journal}{J. Phys. C} \textbf{\bibinfo{volume}{7}},
  \bibinfo{pages}{8383} (\bibinfo{year}{1995}).

\bibitem{Yuan}
\bibinfo{author}{\bibfnamefont{H.~Q.} \bibnamefont{Yuan}},
  \bibinfo{author}{\bibfnamefont{U.}~\bibnamefont{Grimm}},
  \bibinfo{author}{\bibfnamefont{P.}~\bibnamefont{Repetowicz}},
  \bibnamefont{and}
  \bibinfo{author}{\bibfnamefont{M.}~\bibnamefont{Schreiber}},
  \bibinfo{journal}{Phys. Rev. B} \textbf{\bibinfo{volume}{62}},
  \bibinfo{pages}{15569} (\bibinfo{year}{2000}).

\bibitem{Numerique}
\bibinfo{author}{\bibnamefont{{C. Leforestier {\it et~al.}}}},
  \bibinfo{journal}{J. Comp. Phys.} \textbf{\bibinfo{volume}{94}},
  \bibinfo{pages}{59} (\bibinfo{year}{1991}).

\bibitem{KPG}
\bibinfo{author}{\bibfnamefont{R.}~\bibnamefont{Ketzmerick}},
  \bibinfo{author}{\bibfnamefont{G.}~\bibnamefont{Petschel}}, \bibnamefont{and}
  \bibinfo{author}{\bibfnamefont{T.}~\bibnamefont{Geisel}},
  \bibinfo{journal}{Phys. Rev. Lett.} \textbf{\bibinfo{volume}{69}},
  \bibinfo{pages}{695} (\bibinfo{year}{1992}).

\bibitem{Piechon}
\bibinfo{author}{\bibfnamefont{F.}~\bibnamefont{Pi\'echon}},
  \bibinfo{journal}{Phys. Rev. Lett.} \textbf{\bibinfo{volume}{76}},
  \bibinfo{pages}{4375} (\bibinfo{year}{1996}).

\bibitem{Ketzmerick}
\bibinfo{author}{\bibfnamefont{R.}~\bibnamefont{Ketzmerick}},
  \bibinfo{author}{\bibfnamefont{K.}~\bibnamefont{Kruse}},
  \bibinfo{author}{\bibfnamefont{S.}~\bibnamefont{Kraut}}, \bibnamefont{and}
  \bibinfo{author}{\bibfnamefont{T.}~\bibnamefont{Geisel}},
  \bibinfo{journal}{Phys. Rev. Lett.} \textbf{\bibinfo{volume}{79}},
  \bibinfo{pages}{1959} (\bibinfo{year}{1997}).

\bibitem{Behrooz_PRL}
\bibinfo{author}{\bibfnamefont{A.}~\bibnamefont{Behrooz}},
  \bibinfo{author}{\bibnamefont{{M.~J. Burns}}},
  \bibinfo{author}{\bibfnamefont{H.}~\bibnamefont{Deckman}},
  \bibinfo{author}{\bibfnamefont{D.}~\bibnamefont{Levine}},
  \bibinfo{author}{\bibfnamefont{B.}~\bibnamefont{Whitehead}},
  \bibnamefont{and} \bibinfo{author}{\bibnamefont{{P.~M. Chaikin}}},
  \bibinfo{journal}{Phys. Rev. Lett.} \textbf{\bibinfo{volume}{57}},
  \bibinfo{pages}{368} (\bibinfo{year}{1986}).

\bibitem{Behrooz_PRB}
\bibinfo{author}{\bibfnamefont{A.}~\bibnamefont{Behrooz}},
  \bibinfo{author}{\bibnamefont{{M. J. Burns}}},
  \bibinfo{author}{\bibnamefont{D.Levine}},
  \bibinfo{author}{\bibfnamefont{B.}~\bibnamefont{Whitehead}},
  \bibnamefont{and} \bibinfo{author}{\bibnamefont{{P. M. Chaikin}}},
  \bibinfo{journal}{Phys. Rev. B} \textbf{\bibinfo{volume}{35}},
  \bibinfo{pages}{8396} (\bibinfo{year}{1987}).

\bibitem{Wang2}
\bibinfo{author}{\bibfnamefont{Y.~Y.} \bibnamefont{Wang}},
  \bibinfo{author}{\bibfnamefont{R.}~\bibnamefont{Steinmann}},
  \bibinfo{author}{\bibfnamefont{J.}~\bibnamefont{Chaussy}},
  \bibinfo{author}{\bibfnamefont{R.}~\bibnamefont{Rammal}}, \bibnamefont{and}
  \bibinfo{author}{\bibfnamefont{B.}~\bibnamefont{Pannetier}},
  \bibinfo{journal}{Jpn. J. Appl. Phys.} \textbf{\bibinfo{volume}{26}},
  \bibinfo{pages}{1415} (\bibinfo{year}{1987}).

\bibitem{Springer}
\bibinfo{author}{\bibfnamefont{K.}~\bibnamefont{Springer}} \bibnamefont{and}
  \bibinfo{author}{\bibnamefont{{D.~Van Harlingen}}}, \bibinfo{journal}{Phys.
  Rev. B} \textbf{\bibinfo{volume}{36}}, \bibinfo{pages}{7273}
  (\bibinfo{year}{1987}).

\bibitem{Nori1}
\bibinfo{author}{\bibfnamefont{F.}~\bibnamefont{Nori}},
  \bibinfo{author}{\bibfnamefont{Q.}~\bibnamefont{Niu}},
  \bibinfo{author}{\bibfnamefont{E.}~\bibnamefont{Fradkin}}, \bibnamefont{and}
  \bibinfo{author}{\bibfnamefont{S.~J.} \bibnamefont{Chang}},
  \bibinfo{journal}{Phys. Rev. B} \textbf{\bibinfo{volume}{36}},
  \bibinfo{pages}{8338} (\bibinfo{year}{1987}).

\bibitem{Nori3}
\bibinfo{author}{\bibfnamefont{Q.}~\bibnamefont{Niu}} \bibnamefont{and}
  \bibinfo{author}{\bibfnamefont{F.}~\bibnamefont{Nori}},
  \bibinfo{journal}{Phys. Rev. B} \textbf{\bibinfo{volume}{39}},
  \bibinfo{pages}{2134} (\bibinfo{year}{1989}).

\bibitem{Alexander_fils_supra}
\bibinfo{author}{\bibfnamefont{S.}~\bibnamefont{Alexander}},
  \bibinfo{journal}{Phys. Rev. B} \textbf{\bibinfo{volume}{27}},
  \bibinfo{pages}{1541} (\bibinfo{year}{1983}).

\bibitem{DeGennes_fils_supra}
\bibinfo{author}{\bibfnamefont{P.~G.} \bibnamefont{de~Gennes}}
  (\bibinfo{year}{1981}), \bibinfo{note}{{C. R.} Acad. Sci. Ser. B {\bf 292} 9
  and 279}.

\bibitem{Pannetier2}
\bibinfo{author}{\bibfnamefont{B.}~\bibnamefont{Pannetier}},
  \bibinfo{author}{\bibfnamefont{J.}~\bibnamefont{Chaussy}},
  \bibinfo{author}{\bibfnamefont{R.}~\bibnamefont{Rammal}}, \bibnamefont{and}
  \bibinfo{author}{\bibfnamefont{J.~C.} \bibnamefont{Vill\'egier}},
  \bibinfo{journal}{Phys. Rev. Lett.} \textbf{\bibinfo{volume}{53}},
  \bibinfo{pages}{1845} (\bibinfo{year}{1984}).

\bibitem{Abilio_T3}
\bibinfo{author}{\bibfnamefont{C.~C.} \bibnamefont{Abilio}},
  \bibinfo{author}{\bibfnamefont{P.}~\bibnamefont{Butaud}},
  \bibinfo{author}{\bibfnamefont{T.}~\bibnamefont{Fournier}},
  \bibinfo{author}{\bibfnamefont{B.}~\bibnamefont{Pannetier}},
  \bibinfo{author}{\bibfnamefont{J.}~\bibnamefont{Vidal}},
  \bibinfo{author}{\bibfnamefont{S.}~\bibnamefont{Tedesco}}, \bibnamefont{and}
  \bibinfo{author}{\bibfnamefont{B.}~\bibnamefont{Dalzotto}},
  \bibinfo{journal}{Phys. Rev. Lett.} \textbf{\bibinfo{volume}{83}},
  \bibinfo{pages}{5102} (\bibinfo{year}{1999}).

\bibitem{Xiao_Kagome}
\bibinfo{author}{\bibfnamefont{Y.}~\bibnamefont{Xiao}},
  \bibinfo{author}{\bibfnamefont{D.~A.} \bibnamefont{Huse}},
  \bibinfo{author}{\bibfnamefont{P.~M.} \bibnamefont{Chaikin}},
  \bibinfo{author}{\bibfnamefont{M.~J.} \bibnamefont{Higgins}},
  \bibinfo{author}{\bibfnamefont{S.}~\bibnamefont{Bhattacharya}},
  \bibnamefont{and} \bibinfo{author}{\bibfnamefont{D.}~\bibnamefont{Spencer}},
  \bibinfo{journal}{Phys. Rev. B} \textbf{\bibinfo{volume}{65}},
  \bibinfo{pages}{214503} (\bibinfo{year}{2002}).

\bibitem{Lin_Nori}
\bibinfo{author}{\bibfnamefont{Y.~L.} \bibnamefont{Lin}} \bibnamefont{and}
  \bibinfo{author}{\bibfnamefont{F.}~\bibnamefont{Nori}},
  \bibinfo{journal}{Phys. Rev. B} \textbf{\bibinfo{volume}{65}},
  \bibinfo{pages}{214504} (\bibinfo{year}{2002}).

\bibitem{Roche_Review}
\bibinfo{author}{\bibfnamefont{S.}~\bibnamefont{Roche}},
  \bibinfo{author}{\bibfnamefont{D.}~\bibnamefont{Mayou}}, \bibnamefont{and}
  \bibinfo{author}{\bibfnamefont{G.}~\bibnamefont{{Trambly de
  Laissardi\`ere}}}, \bibinfo{journal}{J. Math. Phys.}
  \textbf{\bibinfo{volume}{38}}, \bibinfo{pages}{1794} (\bibinfo{year}{1997}).

\end{thebibliography}
%\bibliographystyle{prsty}

\end{document}